# Mobility Extraction and Quantum Capacitance Impact in High Performance Graphene Field-effect Transistor Devices

Zhihong Chen[1], Joerg Appenzeller[2]

[1] IBM T.J. Watson Research Center, Yorktown Heights, NY 10598, USA, [2] School of Electrical and Computer Engineering and Birck Nanotechnology Center, Purdue University, West Lafayette, IN 47907, USA

## Abstract

The field-effect mobility of graphene devices is discussed. We argue that the graphene ballistic mean free path, $L_{ball}$ can only be extracted by taking into account both, the electrical characteristics and the channel length dependent mobility. In doing so we find a ballistic mean free path of $L_{ball}$=300±100nm at room-temperature for a carrier concentration of ~$10^{12}$cm$^{-2}$ and that a substantial series resistance of around 300Ωμm has to be taken into account. Furthermore, we demonstrate first quantum capacitance measurements on single-layer graphene devices.

## Introduction

With the need to ever further improve the performance of field-effect transistors (FETs) researchers are evaluating various novel channel materials. The most recent candidate that has sparked a substantial amount of excitement is graphene. Graphene occurs to be an ideal choice for FET applications since it combines a) an ultra-thin body for aggressive channel length scaling with b) excellent intrinsic transport properties [1-4] similar to carbon nanotubes but with c) the chance to pattern the desired device structures within a top-down lithographical approach [5,6]..

## Mean Free Path Extraction

First, we will discuss devices that utilize the silicon substrate as a back-gate. This approach reduces the likelihood of multiple-step sample processing - as necessary for a top-gate - impacting the electrical properties of graphene. We have fabricated FETs of various channel lengths (L) and widths (W) using heavily p-doped Si substrates covered with 300nm SiO$_2$, as shown in Fig. 1. Exemplary $I_d$-$V_{gs}$ characteristics of four devices with distinctly different L are shown in Fig. 2. One can clearly identify the expected ambipolar device characteristics with a minimum close to $V_{gs}$=0 that is associated with the Dirac point of graphene. Fig. 3 illustrates schematically how the gate voltage impacts the transport through the graphene layer. While we observe that $I_d(V_{gs})$/W varies from device to device, we find that device characteristics of FETs with L in the one micrometer range and below exhibit almost identical transconductance per channel width ($g_m$/W) (top of Fig. 2). For L>2μm on the other hand, $g_m$/W decreases monotonically with increasing L as evident from the bottom part of Fig. 2. All measurements were performed in the linear $I_d$-$V_{ds}$ range with $V_{ds}$=-10mV (Fig. 4). When employing diffusive transport equations to determine μ, a rather peculiar mobility versus channel length dependence is revealed (dots in Fig. 5). [Note, that our approach is different from other groups' extraction of mobility in that we are using $g_m$ instead of $I_d$/($V_{gs}$-$V_{Dirac}$). Since graphene FETs do not turn off for any gate voltage due to the absence of an energy gap, $I_d$/($V_{gs}$-$V_{Dirac}$) is an inappropriate measure of the device mobility and overestimates μ substantially. In addition, this approach results in a non-physical $V_{gs}$ dependence of μ.] For L larger than around 1μm, μ is constant indicating diffusive transport. However, for smaller channel lengths, μ decreases linearly with decreasing L. The linear dependence of μ on L is simply a result of the fact that $g_m$/W is almost independent of L for constant $C_{ox}$/(WL) and $V_{ds}$. There are two explanations for the observed "artificial" degradation of mobility with decreasing L. 1) μ(L) decreases with L since a substantial series resistance that does not scale with L is present in the devices. 2) μ(L) decreases since the channel length becomes comparable to the scattering length λ and devices exhibit ballistic or quasi-ballistic transport properties. One of the common approaches to extract the mean free path of graphene FETs is by purely relying on the device transfer characteristics and assuming that any conductivity difference from the ballistic limit arises from scattering in the channel. Mean free paths of 10-100nm are frequently extracted in this way [7]. Employing the same approach for our devices indeed results in a ballistic mean free path of around 100nm. However, this value is inconsistent with the obtained µ(L) plot in Fig. 5. A much too steep fit (not shown) is obtained for this λ-value.

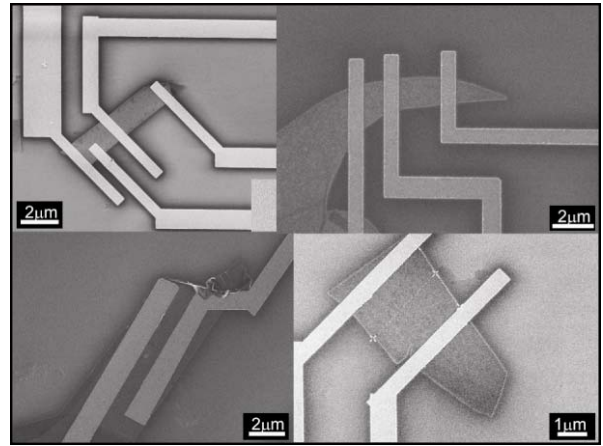

Fig. 1: Scanning electron microscopy images of back-gated graphene field-effect transistors with various channel lengths and widths.

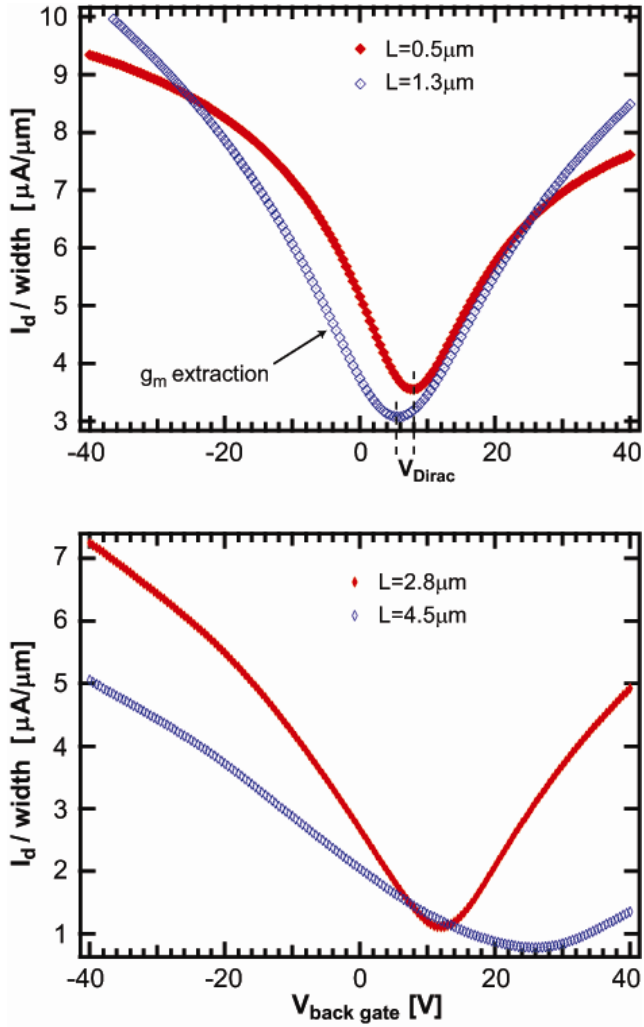

Fig. 2: Room-temperature current per channel width vs. $V_{gs}$ for four samples with different channel length L. $V_{ds}$=-10mV and $t_{ox}$=300nm $SiO_2$.

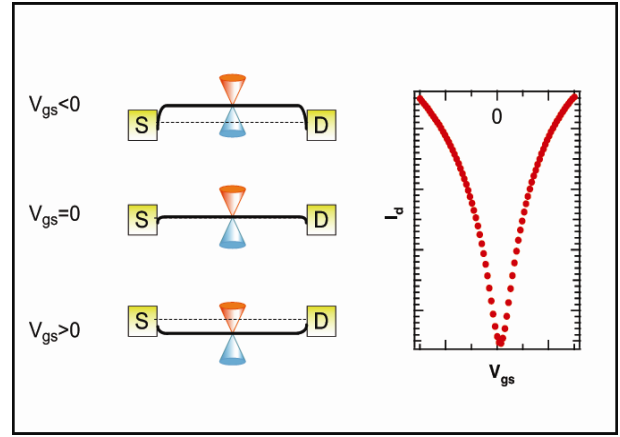

Fig. 3: Schematic of the band movement of graphene as a function of $V_{gs}$. The resulting V-shaped $I_d$-$V_{gs}$ is shown on the right. High currents are obtained for both positive and negative gate voltages.

Instead, using exclusively the data from Fig. 5 would result in a ballistic mean free path of around 600nm when employing µ=µ$_0$*L/(λ+L). The apparent inconsistency between the two methods can be overcome and a coherent picture can be obtained if both aspects 1) and 2) from above are considered simultaneously. The black curve in Fig. 5 considers both, the transition into the ballistic regime as well as a length independent series resistance. We extract a mean free path of 300±100nm and a series resistance of around 300Ωµm (R*W) that can explain our $I_d$-$V_{gs}$ and µ(L) data sets consistently. Such an additional resistance is not as surprising as it appears at a first glance. In fact, an intrinsic, graphene related series resistance has been previously observed and was explained by the presence of an internal p/n junction inside the graphene and by contact induced scattering at the metal/graphene interface [8]. Here we show that extracting the intrinsic graphene properties requires considering this series resistance and that it is important to design devices accordingly. Note, that our statement about the ballistic mean free path value is **only valid** for high enough carrier concentrations ($n_s$) since $g_m$ is extracted in the linear $I_d$-$V_{gs}$ regime at voltages that correspond to $n_s$-values of around $10^{12}$cm$^{-2}$ and above.

## Quantum Capacitance Measurements

Next, we turn our attention to the device capacitance. In order to verify that the calculated $C_{ox}$ values used to extract µ are accurate and to elucidate on the impact of scaling, we have performed C-V measurements on various top-gated graphene FETs. The device layout as well as an SEM image of a representative device is shown in Fig. 6. Considering the conducting graphene layer without a band gap as the counter electrode to the metal top gate, one may expect to measure a constant capacitance $C_{ox}$. However, as illustrated in the capacitor network in Fig. 6, the so-called quantum capacitance needs to be taken into account for low-dimensional systems in general. The quantum capacitance $C_q$ describes the response of the charge inside the channel to the conduction and valence band movement.

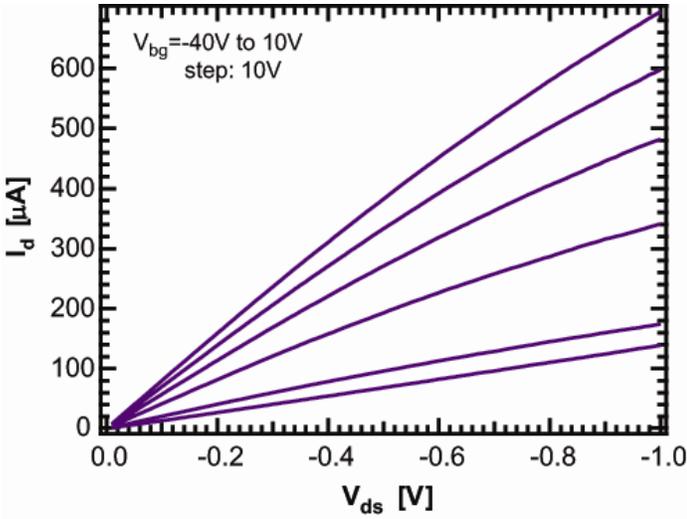

Fig. 4: $I_d$-$V_{ds}$ characteristics of a back-gated graphene FET at room-temperature.

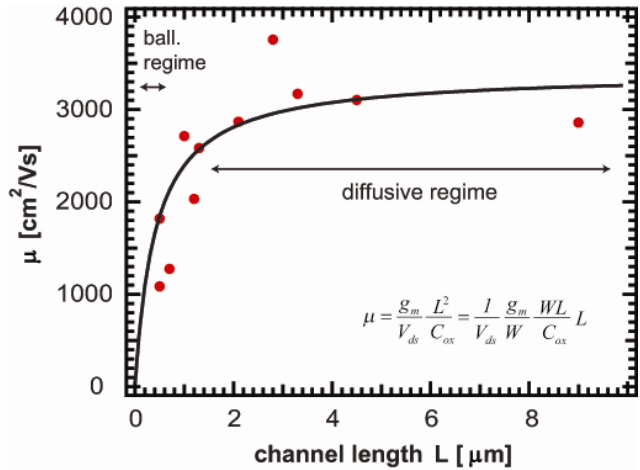

Fig. 5: Mobility versus channel length extracted for various graphene FETs assuming diffusive transport conditions.

$C_q$ is proportional to the density of states (DOS). While $C_q$ is large in conventional systems and can be neglected accordingly, it can be very small in low-dimensional devices in this way becoming the dominant contribution to the total capacitance $C_{tot}$.

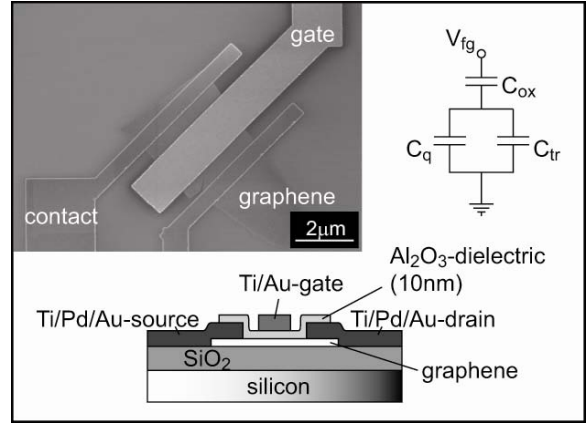

Fig. 6: Bottom: Schematic layout of a top-gated graphene FET. Top: Partially-gated device (left) and capacitance components involved (right).

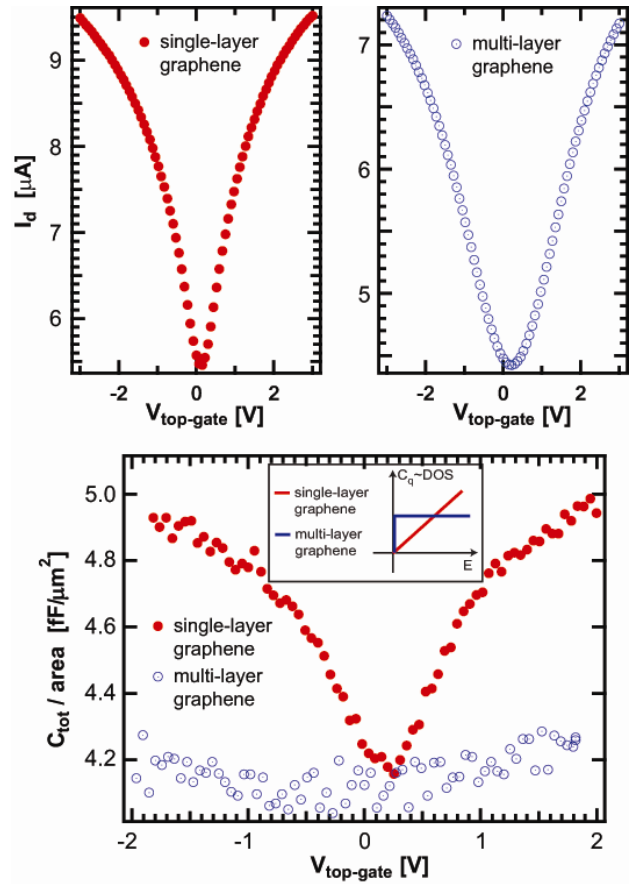

Fig. 7: $I_d$-$V_{gs}$ (top) and total capacitance (bottom) for a top-gated single layer and multi-layer graphene FET with 10nm $Al_2O_3$.

Fig. 7 shows $I_d$-$V_{gs}$ and $C_{tot}$-$V_{gs}$ measurements for a single- and a multi-layer graphene FET. While both devices show a clear variation of current with gate voltage independent of the number of graphene layers, $C_q$ only shows a pronounced $V_{gs}$-dependence for the single layer graphene FET. As illustrated in the inset of the bottom graph of Fig. 7, the difference in DOS (~$C_q$) is the key to understand this behavior, a distinguishing fact that had not been noticed previously. Next we extract $C_q(E)$ from the measurements using the capacitor network in Fig. 6. For a constant oxide and trap capacitance of $C_{ox}$=5.6fF/µm$^2$ and $C_{tr}$=10fF/µm$^2$ respectively the anticipated $C_q(E)$-dependence is obtained (Fig. 8). Not only is the extracted $C_{ox}$-value close to the geometrical value of 5.8fF/µm$^2$, but the actual $C_q$-values as a function of energy are the same for different single-layer graphene devices and are following the theoretical expectation when assuming a density of states $DOS = aE/\hbar^2 v_F^2 \pi$ with $a$=1. Note, that this DOS is a quarter of the usually found expression. While the high ac voltage of 0.2V employed to perform the capacitance measurement can explain a reduction of $C_q$ by a factor of two [9], it is currently unclear why our experimental data are off by another factor of 2. We note that the conversion of the gate voltage axis into an energy axis is a possible source of error in this context.

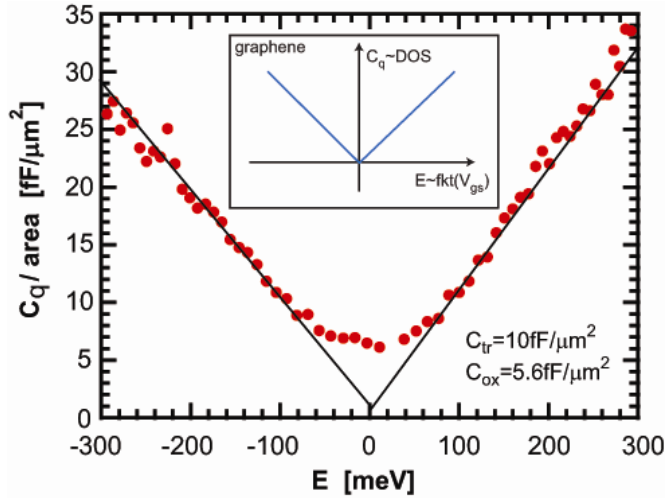

Fig. 8: Extracted quantum capacitance $C_q$ as a function of energy. For E=0 a vanishing quantum capacitance is expected at the Dirac point. The finite measured values around E=0 are a result of the impact from "electron-hole puddles" in the substrate.

## Conclusion

In conclusion, we measured mobility vs. channel length for back-gated graphene FETs and identified the existence of a series resistance in the vicinity of the graphene contact area. Using both, the $I_d$-$V_{gs}$ characteristics and µ(L) data, we are able to extract the mean free path of the graphene channel to be 300±100nm for graphene FETs fabricated on SiO$_2$ substrates. Next, we showed first capacitance measurements on top-gated FETs and extracted the quantum capacitance of graphene. Our experimental findings give critical insights into the importance of $C_q$ for scaled graphene FETs. Even in the case of $t_{ox}$=10nm as in the present case, the small $C_q$-value close to E=0 dominates the total device capacitance and needs to be included in the mobility extraction for the corresponding carrier concentrations. More aggressively scaled graphene FETs require taking into account an even larger energy - and correspondingly carrier concentration - range. Our results also indicate that for the thick oxides used in the back-gated devices we have analyzed above, it is justified to ignore $C_q$ at the gate voltage levels used to extract µ.

## Acknowledgement

The authors would like to thank Dr. Oki Gunawan for his help with the C-V measurement set-up.